\documentclass[aps,prl,twocolumn,superscriptaddress,showpacs]{revtex4-1}
\usepackage{epsfig}
\usepackage{amsmath}
\usepackage{amssymb}

\newcommand{\sigmau}{$e\times 10^{12}/\mathrm{cm}^2$}

\begin{document}

\title{Spontaneous polarization induced doping in quasi-free standing epitaxial
graphene on silicon carbide from density functional theory}
%\title{Accurate determination (probing?) of doping in perfect and defected quasi-free standing epitaxial graphene on silicon carbides from density functional theory}
%\title{First-principles evaluation of {\it p}-type doping level in quasi-free standing epitaxial graphene on H-passivated SiC(0001) surfaces}

\author{J. S\l awi\'{n}ska}
\affiliation{Department of Solid State Physics, University of \L \'{o}d\'{z}, Pomorska 149/153, 90236 \L \'{o}d\'{z}, Poland}
\affiliation{Instituto de Ciencia de Materiales de Madrid, ICMM-CSIC, Cantoblanco, 28049 Madrid, Spain.}

\author{H. Aramberri}
\affiliation{Instituto de Ciencia de Materiales de Madrid, ICMM-CSIC, Cantoblanco, 28049 Madrid, Spain.}

\author{J. I. Cerd\'a}
\affiliation{Instituto de Ciencia de Materiales de Madrid, ICMM-CSIC, Cantoblanco, 28049 Madrid, Spain.}

\date{\today}
\begin{abstract}
By means of density functional theory (DFT) calculations we have quantitatively
estimated the impact of the spontaneous polarization (SP) of the SiC(0001)
substrate on the electronic properties of quasi-freestanding graphene (QFG)
decoupled from the SiC by H intercalation. 
To correctly include within standard DFT slab calculations
the influence of the SP, which is a bulk property, on a surface
confined property such as the graphene's doping,
we attach a double gold layer at the C-terminated bottom of the 
slab which introduces a metal induced gap state that pins the chemical potential
within the gap. Furthermore, expanding the interlayer distances at the 
bottom of the slab creates a local dipole moment which counters that
arising from the slab's polar character and allows to align
the location of the graphene's Dirac point (DP) for
cubic SiC(111) with the chemical potential. Thus, the DP shifts
obtained for other polytypes under the same slab model
become an almost direct measurement of the SP-induced doping. 
Our results confirm the recent proposal that the SP induces the experimentally 
observed $p$-type doping in the graphene layer which can achieve DP shifts of
up to several hundreds of meV (or equivalently, $\sim 10^{13} e$/cm$^2$) for 
specific polytypes. The doping is found to increase with the hexagonality of
the polytype and its thickness. For the slab thickensses considered 
(6-12 SiC bilayers) an ample, almost continuous, range of doping 
values can be achieved by tuning the number of stacking defects and
their location with respect to the surface. 
The slab model is next generalized by performing large scale DFT calculations
where self-doping is included in the QFG via
point defects (vacancy plus a H atom) thus allowing to estimate 
the interplay between both sources of $p$-doping (SP- versus 
defect-induced) which turns out to be essentially additive.
\end{abstract}
\maketitle

{\it Introduction--} Epitaxial graphene (EG) on silicon carbide (SiC) has been demonstrated to be an
excellent material for high-performance technological applications. \cite{hft1, hft2, hft3, frequency-mixers, resistance} On the other hand, quasi-freestanding
graphene (QFG) on SiC obtained from EG by hydrogen intercalation holds an even 
greater promise for the realization of graphene-based electronic devices 
because of its facile large-scale production together with excellent transport 
characteristics~\cite{raman, goler, transistors, riedl}. 
The latter mainly arise from the efficient reduction of the interaction between
graphene (G) and SiC, otherwise strongly coupled, by the intercalated H layer. 
Although this weak interaction fully preserves the Dirac 
cones~\cite{riedl, cubic, fortinew, ourcarbon}, a $p$-type doping is routinely 
detected in experiments~\cite{riedl,raman,jjap,forti,virojanadara, newrajput} 
which can be as large as 
%\textit{320~meV as  observed by Rajput~{\it et al}~\cite{newrajput} for QFG samples on 6H-SiC(0001) or $\sim1.5\times 10^{13}$ cm$^{-2}$ as recently reported on 4H-SiC(0001) by Urban~{\it et al}~\cite{nitro} while for cubic 3C-SiC(111) Coletti~{\it et al} found a slight $n$-type doping~\cite{cubic,newtrilayer}. Despite these results, among several others~\cite{transistors, goler, elpho, newexp, schottky}, suggest certain relationship between the QFG doping and the SiC polytype, its ultimate origin has not yet been convincingly established~\cite{riedl,virojanadara,schottky,transistors,jjap,newtrilayer} and remains controversial even in the most recent QFG-related works~\cite{newrajput, newpasquarello, newth, fortinew, nitro}.}
$\sim5.5\times 10^{12}$ cm$^{-2}$ as observed by Speck~{\it et al}~\cite{raman}
for QFG samples on 6H-SiC(0001) or even $\sim2.0\times 10^{13}$ cm$^{-2}$ as 
recently reported on 4H-SiC(0001) by Urban~{\it et al}~\cite{nitro}, while for 
cubic 3C-SiC(111) Coletti~{\it et al} found a slight $n$-type 
doping~\cite{cubic,newtrilayer}. 

Based on these results, among several others~\cite{transistors, goler, elpho, newexp, schottky},
Ristein {\it et al}~\cite{pdope} pointed to a relationship between the G doping and
the hexagonality of the SiC polytype employed as substrate, and proposed
a {\it macroscopic} spontaneous polarization (SP) doping model whereby the 
substrate's SP creates a pseudo-charge at the surface equivalent to a real 
acceptor layer. The model explains the sign of the doping
(since the Si-terminated surface exhibits a negative SP for all polytypes) as 
well as its dependence on the hexagonality of the polytype, and has found 
further support in several works~\cite{elpho,newtrilayer,fortinew,davydov}.
Mammadov {\it et al}~\cite{newrms} have very recently corroborated such
model by means of a systematic angle-resolved photoemission electron 
spectroscopy (ARPES) study for 3C-, 4H- and 6H-SiC QFG-systems.
Two sources
of doping are identified in this study: a surface band bending arising from
the bulk dopants and the SP. The former would be responsible for the mild
$n$-type doping encountered in 3C-SiC(111) samples~\cite{cubic,newtrilayer,newrms}
while the latter would account for the $\sim1.5$ factor between
hole dopings in 4H- and 6H-SiC samples~\cite{newrms}. Although not addressed
in that work, a third main source is the self-doping induced by the presence
of intrinsic defects in the QFG (vacancies and/or adatoms), thoroughly studied 
within theory~\cite{ourcarbon, yazyev} but not so well
experimentally \cite{raman,concentrations,berger,bufferlayer,epjb}.

In this Letter we investigate the relationship between the SP and the graphene's $p$-type doping in perfect as well as defected QFG for several H-passivated hexagonal SiC polytypes via density functional theory (DFT) based calculations. The importance of employing a first principles approach to this end is manifold: it represents a powerful predictive tool when tuning the surface density of $e$/hole carriers in G-based devices~\cite{schottky, hft1, hft2, hft3, frequency-mixers, resistance}, provides a unique way to quantify on equal footings the competition between self-doping and SP and, not the least, allows to explore the validity of the macroscopic dielectric theory when the system size is
shrinked to the nanoscale, as it is not obvious if a direct relationship 
between the SP and the doping charge in the QFG will still hold.
Nevertheless, first principles slab calculations --aiming to model a semi-infinite surface-- have not yet been attempted due to at least three non-trivial 
issues which need to be resolved:
(i) combining a bulk and a surface confined property (SP and G's doping, 
respectively) typically requires rather thick slabs to achieve convergence
(this is specially true for dielectrics with long screening lengths), 
(ii) an appropiate boundary condition at the bottom of the slab which pins the 
chemical potential, $\mu$, within the gap regardless of the selected polytype
is a prerequisite to make any differences in the doping 
among them meaningful and, (iii) the polar character of 
SiC(111)/(0001) oriented slabs introduces an additional electric field across 
the dielectric which may considerably alter the 
final doping level. Additionally, the large dispersion of the $\pi$ bands
forming the Dirac cones requires a hyperfine sampling of the Brillouin Zone 
(BZ) for an accurate estimation of the doping charge (see below), 
thus increasing considerably the 
computational time. Paradoxically, the associated low G-projected density of 
states (PDOS) around the DP makes the DP shift with respect to $\mu$,
$\Delta$DP, a highly precise gauge for the G doping, specially from the
experimental side.

The first goal of this work is therefore to set up a calculation strategy that 
overcomes the above fundamental drawbacks. 
After considering different slab models we have
found that a G/H/(SiC)$_n$/Au$_2$ slab with
a double gold layer attached to the lower C-dangling bonds meets satisfactorily
the above requirements. The model is then employed to calculate the QFG dopings
of different polytypes
as a function of the number of SiC bilayers (BLs) in the slab, $n$.
Apart from the cubic 3C-SiC(111), we have considered 
those with largest hexagonality, namely, 2H-, 4H- and 6H-SiC(0001), having
a stacking defect (SD) every two, three and four BLs, respectively. 
The combined effect of SP and defect-induced self-doping will be addressed at
the end of this letter.

\begin{figure}[!hbt]
\includegraphics[scale=0.4]{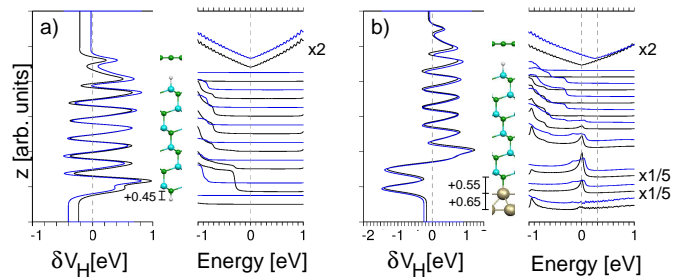}
\caption{ (a) left: $\delta V_H(z)$ profile along a G/H/(SiC)$_6$/H slab depicted at the center and, right: the corresponding DOS projected, in ascending order, on the bottom H layer, the six SiC bilayers, the intercalated H layer and the G; the bonds between C-terminated bottom of the slab and H atoms have been elongated by 0.45 \AA\, with respect to their optimized values (1.56 vs 1.11 \AA). Black and blue lines correspond to 3C-SiC(111) and 2H-SiC(0001), respectively. (b) Same as in (a), but a G/H/(SiC)$_6$/Au$_2$ slab with gold instead of H termination is shown; the C-Au and Au-Au bonds have been elongated by 0.55\AA\, and 0.65\AA\, respectively. The depicted slab geometries in each figure correspond to 
the 3C-SiC(111) case.
         \label{au2}}
\end{figure}

{\it Methods--} All presented calculations have been performed with the GREEN code~\cite{green} and its interface to the DFT based SIESTA package~\cite{siesta}. We employed the generalized gradient approximation \cite{gga} and included 
semi-empirical van der Waals interactions~\cite{vdw} to account for the weak 
G/H/SiC interaction. Dipole-dipole interactions between spurious slab replicas
were removed via the usual dipole-dipole corrections~\cite{dipole} (DDC) thus 
leading to vanishing electric fields in the vacuum.
The rest of the calculation parameters are described in detail
in Ref.~\cite{prb}. The G/H/(SiC)$_n$/Au$_2$ model slabs comprised 
a ($2\times2$) graphene layer placed on top of $n$ $(\sqrt{3}\times\sqrt{3})R30^\circ$
SiC(111)/(0001) bilayers (BLs) with each Si dangling bond at the upper surface 
saturated by a hydrogen atom, while the bottom C dangling bonds were saturated
by placing a gold plane in registry with the carbon atoms and a second 
one following an $fcc$ stacking sequence (see inset in Figure~\ref{au2}(b)). 
The systems were relaxed freezing the
inner SiC BLs to bulk-like positions optimized independently for each 
polytype. In order to obtain accurate doping charges, large (100$\times$100)
$k$-supercells were employed to sample the Brillouin zone~\cite{prb}.

{\it SP induced doping--}
Let us first address the problems outlined above related to the use of a slab 
geometry when trying to estimate the G doping. For the most common practice
of capping the lower C-terminated surface with another hydrogen layer these
shortcomings become patent. In Figure~\ref{au2}(a) we plot the 2D averaged
Hartree potential profiles, $V_H(z)$, and the layer resolved PDOS for a 
G/H/(SiC)$_6$/H slab assuming a 3C- (black lines) or 2H-SiC (blue) 
stacking. Since considerable dipole moments were found in both cases
($\Delta V=$0.8 and 1.5~eV for the 3C and 2H slabs, respectively) 
we elongated the C-H bonds with respect to their optimized values 
(see Fig.~\ref{au2}(a)) thus reducing the potential drops by as much as 0.7~eV.
Unfortunately, and regardless of this expansion, the DP remains pinned at 
the chemical potential for both slabs. The reason is the absence of
any gap states at the bottom of the slab, so that charge neutrality at the 
graphene layer locks $\mu$ at the DP.
A natural way to overcome this drawback is to introduce
states within the gap and localized at the bottom part of the slab. If
their associated DOS is much larger than that of the G, then they
should pin the slab's chemical potential just as bulk dopants/defects fix 
$\mu$ in a real surface, leaving the G's Dirac cones {\it free} to 
trap or release electrons in order to screen any internal fields. Attaching
a double gold layer at the C-ended lower layer does indeed yield the desired 
boundary condition. In Fig.~\ref{au2}(b) we present the potential profiles 
and PDOS of such a G/H/(SiC)$_6$/Au$_{2}$ slab for 3C- and 2H-SiC.
The potential drops ($\Delta V$=0.3 and 0.5~eV) were reduced by 0.2~eV after
expanding the Au-C and Au-Au interlayer distances by large amounts ($>$0.5~\AA).
We stress that although the expanded bottom geometry is not realistic, its
purpose is to modify the local dipole at the bottom of the slab in order to
reduce the total slab's dipole moment.
Inspection of the PDOS show that metal induced gap states (MIGSs) appear in both
cases as a large broad peak mainly localized at the interface between the last 
SiC BL and the gold plane, penetrating around three BLs into the substrate. The
MIGSs now determine the slab's chemical potential and, notably,
the doping in the QFG layer has almost vanished for the 3C case
(black line), while the 2H slab (blue) displays a clear $p$-doping with a 
considerable DP shift of around 250~meV.
We have also considered attaching single and triple gold layers at the bottom
of the slab to find that the former (latter) yields a slight $n$ ($p$)-type 
doping in the G~\cite{prb}. Such slabs may therefore be employed as model
systems where bulk dopants already induce certain level of doping~\cite{newrms}.

\begin{figure}[!hbt]
\includegraphics[width=\columnwidth]{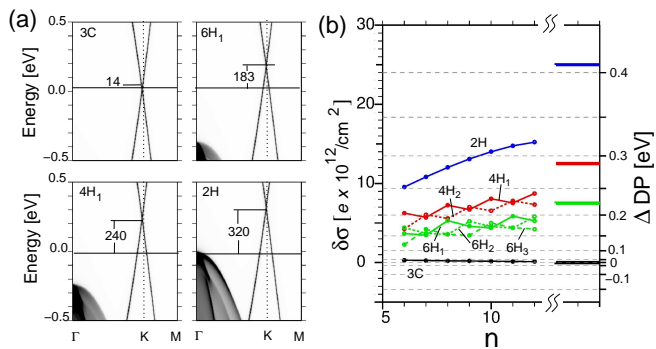}
\caption{(a) DOS($k, E$) projected on G, H and first three SiC BLs of a G/H/SiC
         semi-infinite surface calculated after matching the Green's function of
         a G/H/(SiC)$_{12}$/Au$_2$ slab with that of the corresponding SiC bulk.
         Four different SiC polytype surfaces are shown: 3C, 6H$_1$, 4H$_1$ and
         2H. $\Delta$DP is indicated in each plot in meV. 
	 (b) Doping of the graphene layer for all G/(SiC)$_n$/Au$_2$ slabs 
         considered in this work as a function of $n$, the SiC politype and the
         location of the SD closest to the surface. Left axis gives the 
         doping surface charge density, $\delta\sigma$, and right axis the 
         Dirac point shift, $\Delta$DP=DP$-\mu$ (non-linear scale).
         Thick horizontal lines at 
         the right indicate the bulk SP associated to each polytype.
                  \label{dopei}}
\end{figure}

The above analysis provides clear evidence that our slab model allows to 
address the impact that the SDs present in the film have on the graphene layer.
Figure~\ref{dopei}(a) shows a precise picture of the Dirac cones in the form of
PDOS($k,E$) for a 3C, 6H, 4H, and 2H substrate. They have been calculated for
a semi-infinite geometry after replacing the Hamiltonian of the lower half of 
the G/H/(SiC)$_{12}$/Au$_2$ slab by that of the corresponding bulk 
polytype~\cite{loit}\footnote{The semi-infinite approach does not 
modify the DP shifts deduced from the slab calculations since the 
bulk employed in the matching process does not contribute to the SP. See also
Ref.\cite{prb}}.
Apart from the negligible value of 14~meV for cubic 3C-SiC(111), the graphs clearly reveal how the DP shifts increase with the level of hexagonality attaining several hundreds of meV.

We now turn our attention on how these dopings evolve with the film thickness.
Figure~\ref{dopei}(b) 
illustrates the dependence of the graphene doping charge
$\delta\sigma$ (left axis) and $\Delta$DP (right axis) as a function of the slab
thickness and the polytype considered. For the 6H and 4H cases we have defined 
different subsets of data depending on the location of the SD closest to
the surface, which may be either between the first and second BLs (4H$_1$ and 
6H$_1$) or the second and third (4H$_2$ and 6H$_2$) or the third and fourth 
(6H$_3$).  Except for the 3C case, which remains with a negligible doping for 
all thicknesses, the rest of plots reveal an overall increase of $\delta\sigma$
with $n$ approaching their respective bulk SP values (indicated by thick
horizontal lines at the right of the plot).
The 6H and 4H polytypes show a stair-like behavior which can be 
understood by noting that as $n$ is increased by one, the added BL may or may 
not introduce an additional SD in the slab. In the former case, the slab's SP 
increases and $\delta\sigma$ shows an abrupt raise, whereas in the latter case,
the crystalline region increases and hence, the depolarization is more efficient
leading to a slight decrease of the doping. For the 2H case, on the other hand,
$\delta\sigma$ shows a smooth behaviour
with no jumps due to the absence of crystalline regions. However,
the SP for this polytype is so large that we find an early saturation of 
$\delta\sigma$ due to the crossing of the valence band maximum (VBM) by the 
chemical potential at $n=12$ (see Fig.~\ref{dopei}(a)), so that for larger 
thicknesses the SiC VBs will also contribute to the screening of the SP.

The QFG doping dependence displayed in Fig.~\ref{dopei}
constitutes one of the central results of this work as it clearly establishes
a relationship between the SP and QFG $p$-doping also providing an estimate
for the slab size required to reach the bulk SP limit;
at the largest thickness considered, $n=12$, the doping 
amounts to 60-80\% of the SP while, from extrapolation, we expect that almost
a 100\% should be already reached at $n\gtrsim 20$~BLs.
Simple electrostatic arguments dictate that the
difference between $\delta\sigma$ and the SP can be entirely attributed to the 
macroscopic electric field within the dielectric, $\overline{E}$ (generally 
known as the depolarization field) via~\cite{prb}: 
$\epsilon_0 \epsilon_r \overline{E} = \delta\sigma + SP$. 

A further remarkable issue in Fig.~\ref{dopei}(b)
is the fact that an ample range of dopings can be obtained in an almost
continuous way by controlling the number of SDs, their density and their
proximity to the surface G layer. This is 
in line with the wide spread of experimental values 
reported for the graphene's doping within the same polytype. For instance, for 
the most commonly used 6H-SiC(0001) substrate, our calculated dopings fall
in the $4-6$~\sigmau\ range, in excellent agreement with a number of 
experimental results varying from 2.0 to 6.2 $\times 10^{12}$ cm$^{-2}$~\cite{newrms, raman, elpho, schottky}. Smaller $\delta \sigma$ values ($\sim2$~\sigmau)
have also been reported~\cite{transistors,riedl,forti} which, in light of our 
results, might be associated with defects or impurities/dopants either at the G
layer or in the bulk~\cite{newrms}, or even to a cubic termination at the
surface~\cite{forti}. On the other hand, few values of $\delta \sigma$
have been reported for a 4H-SiC(0001) substrate: 
Mammadov {\it et al}~\cite{newrms} obtained 6.9 and 8.6~\sigmau\ for an
$n$-type doped and a semi-insulating substrate, respectively, which 
is in very good accordance with our values for the 4H$_{1/2}$ at $n\gtrsim12$,
while the Hall measurements of Urban~{\it et al}~\cite{nitro} yielded 
charges of $15-20$~\sigmau\ which signifcantly exceed the bulk SP, again 
suggesting the presence of other sources of doping. Unfortunately, we are not
aware of any QFG experiments carried out on a 2H surface.

\begin{figure}
\includegraphics{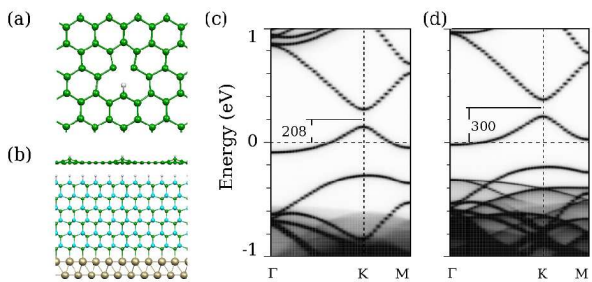}
\caption{Top (a) and side (b) view of the geometry corresponding to a vacancy 
embedded in an ($8\times8$) supercell of G/H/2H-SiC(0001)/Au$_{2}$. In (a) only
graphene atoms and the H atom saturating the dangling bond are shown. 
DOS($\vec{k},E$) projected on the first layers of the surface for 3C-SiC(111) 
(c) and 2H-SiC(0001) (d). Doping levels are indicated in meV.
         \label{vacancy} }
\end{figure}

{\it Competition between SP and self-doping--} Finally, and in order to reach 
a general picture of the doping in QFG systems, we incorporate defects
in the G layer within our gold terminated slab model.
Among the various types of point defects studied in the literature~\cite{yazyev, ourcarbon, defects} we 
consider the Jahn-Teller distorted vacancy structure where two
C dangling bonds establish a bond among them inducing a pentagonal
structure, while the third C dangling bond is saturated by a H atom
(see Figure~\ref{vacancy}(a)). We make this choice because 
this defect structure hardly alters the Dirac cones and, at the
same time, yields a considerable $p$-type doping~\cite{ourcarbon}. Due to
computational limitations, the defect is embedded in an (8$\times$8) G supercell
corresponding to a somewhat large concentration of 0.8$\%$ (typical experimental
values obtained for EG samples are of the order of 0.006$\%$), while the number
of SiC BLs was set to $n$=6, leading to a total of 848 atoms in the unit cell
(see Fig.~\ref{vacancy}(b)).

PDOS($\vec{k},E$) maps are shown in Figs.~\ref{vacancy}(c) and (d), for the
3C and 2H polytypes, respectively. Although a band gap of $\sim$150~meV appears
between the Dirac bands in both cases due to the large defect concentration,
it is still possible to estimate their doping levels.
For the 3C cubic case, a considerable hole concentration of 20.2~\sigmau\
($\Delta$DP=208~meV) is induced in the $\pi$ bands in order to compensate the 
electron charge accumulated in the vicinity of the defect~\cite{ourcarbon}. 
As expected, the doping is even larger for the 2H slab, $\Delta$DP=300~meV
or 31.3~\sigmau, since the SP further contributes to the doping. 
Most importantly, the difference in $\delta\sigma$
between both polytypes is 11~\sigmau, which coincides very nicely
with the SP-induced doping of 10~\sigmau\ found for undefected 2H slabs at
$n=6$ (blue line in Fig.~\ref{dopei}(b)) indicating
that the SP- and self-doping mechanisms are basically additive. This is not
a trivial result as it is not clear {\it a priori} if the self-doping
mechanism will affect the depolarization field $\overline{E}$ which, as mentioned
above, plays an important role in the final G doping.
Our results thus suggest that the self-doping, already very large
on its own, hardly alters the incomplete compensation of the SP pseudo-charge
by the G's $\pi$ bands.

In summary, we have presented a methodology within the framework of standard
DFT slab calculations that accounts for the contribution of the SiC substrate's
SP to the QFG doping. The scheme relies on the pinning of the slab's chemical
potential at the bottom of the slab via the creation of MIGSs states after
saturating the lower C-dangling bonds with a gold bilayer. One may additionally
incorporate self-doping contributions due to defects or adsorbates in the G
layer. Application to QFG on 6H-, 4H- and 2H-SiC(0001) substrates indicates
that full compensation of the SP by the G doping should occur at
thicknesses of $n\gtrsim 20$~BLs, while thinner slabs yield an ample range
of dopings depending on the polytype and the precise termination of the
surface. Other sources of doping, such as self-doping in the G layer, or
bulk dopants, may also be incorporated into the calculations.
Apart from the obvious applicability of the
analysis to ultrathin SiC films, the scheme should also work satisfactorily
in other dielectrics exhibiting an SP although the nature of the metallic
layer and the interlayer expansions required to minimize the surface dipole
will in general need to be tuned for the specific system.

\begin{acknowledgements}
J.S. acknowledges Polish Ministry of Science and Higher Education for financing
the postdoctoral stay at the ICMM-CSIC in the frame of the fellowship Mobility 
Plus. H.A. and J.C. acknowledge financial support from the Spanish Ministry of 
Innovation and Science under contract Nos.~MAT2012-38045-C04-04 
and MAT2013-47878-C2-R respectively.
\end{acknowledgements}

%\bibliography{letter}

\end{document}